\documentclass{elsart}
\usepackage{epsfig}
\usepackage{graphicx}
\usepackage[mathscr]{eucal}

\begin{document}

\begin{frontmatter}
\title{On the nature of cosmic rays above the Greisen--Zatsepin--Kuz'min 
cut off}
\author{L. A. Anchordoqui, M. Kirasirova, T. P. McCauley,} 
\author{S. Reucroft, and J. D. Swain}  
\address{Department of Physics, Northeastern University, Boston, MA 02115}

\begin{abstract}

A re-examination of the atmospheric cascade profile of the highest 
energy cosmic ray is presented. The study includes  air-shower 
simulations considering different cross sections, particle multiplicity 
and variation of the hadronic-event-generator to model interactions 
above 200 GeV. The analysis provides evidence that a medium mass 
nucleus primary reproduces the shower profile quite well. This 
result does not support the idea, 
increasingly popular at present, that the highest energy particles are 
protons, derived from the decay of supermassive relic particles.
On the other hand, we show that  debris of relativistic 
super-heavy nuclei, which can survive a 100 Mpc journey 
through the primeval radiation are likely to generate such a kind of cascade.

\noindent {\it PACS number(s):} 96.40.Tv, 13.85.Tp, 96.40Pq.
\end{abstract}
\end{frontmatter}
\newpage

Shortly after the microwave background radiation (MBR) was 
discovered \cite{mbr}, it became clear that ultra high energy 
cosmic rays (CRs) undergo 
reactions with the relic photons yielding a steep drop in their energy 
attenuation length. Therefore, if the CR sources are all at cosmological 
distances, the energy spectrum would exhibit a Greisen-Zatsepin-Kuz'min (GZK) 
cut off slightly above 100 EeV (1 EeV $\equiv 10^{18}$eV) \cite{gzk}.
The spectral cut off is less sharp for nearby sources, {\it viz.}, distances 
$< 100$ Mpc. 

Over the last few years, several giant air showers have been detected, which 
imply the  arrival of ``super--GZK'' particles \cite{yd}. 
Namely, the Haverah Park experiment reported several events with energies 
near or slightly above 100 EeV \cite{hp}. On October 15, 1991 the 
Fly's Eye experiment recorded the highest energy cosmic ray ever detected on 
Earth ($E \sim 300$ EeV) \cite{FE}.\footnote{Near the arrival 
direction of this event the Yakutsk experiment detected an extensive air 
shower whose energy may also reach 300 EeV \cite{yakutsk}.} More recently, 
the Akeno Giant Air Shower Array (AGASA) has reported several remarkable 
CR events scattered across half the sky dramatically confirming that 
the CR spectrum does not end with the GZK cut off \cite{agasa}.

Many theories have been suggested to explain the origin and nature 
of these particles. At the most general level, the production 
mechanisms 
can be classified into  two distinct groups: ``bottom-up'' acceleration arena, 
and ``top-down'' decay scenario, each of them branching out into various 
classes.  

The ``bottom-up'' mechanism usually implies the stretching of rather 
well known acceleration processes to their theoretical 
limits (and sometimes beyond). 
It involves particle acceleration in the 
accretion flows of cosmological structures. Examples are one-shot 
acceleration in active galactic nuclei (AGNs) \cite{prot}, large scale shocks
resulting from structure formation \cite{nor}, acceleration in quasar 
remnant cores of nearby giant elliptical galaxies \cite{quasars}, 
collisions of galaxies \cite{ces}, acceleration in catastrophic events
associated with gamma ray bursts \cite{wax},  
and acceleration in hot spots of extragalactic radio sources \cite{rabir}.

On the other hand, the ``top-down'' scenario (mischievously) escapes from the 
acceleration problem by assuming that charged and neutral primaries simply 
arise by quantum mechanical decay of supermassive elementary $X$ particles
($m_X \sim 10^{25} - 10^{28}$ eV) \cite{BS}. Sources of these exotic 
particles at present could be topological defects (TDs) left over from early 
universe phase transitions associated with spontaneous symmetry breaking 
underlying unified models of high energy interactions. Due to their 
topological stability, the TDs (magnetic monopoles, 
cosmic strings, domain walls, etc.) can survive forever with $X$ particles 
trapped inside them. Nevertheless, from time to time, some TDs can be 
destroyed through collapse, annihilation, or other processes, and the 
energy stored is released in the form of massive quanta 
that would typically decay into quarks and leptons. The subsequent 
hadronization of the quarks would produce jets of hadrons containing mainly 
light mesons (pions) together with a small fraction (3\%) of nucleons. 
Thus, the ``top-down'' injection spectra would  be dominated by 
gamma rays and neutrinos originated by  pion decays. 

There have been arguments for some time that the cosmic 
photon background makes the universe opaque to ultra high energy 
gamma-rays \cite{gamma}. The most efficient target for these rays to  
produce $e^+ e^-$ pairs are background photons of 
energy $\sim m_e^2 / E \leq 10^{-6}$eV $\approx$ 100 MHz. The mean free 
path for a  100 EeV (1000 EeV) photon to annihilate on the radio background 
is believed to be  10-40 Mpc (1-5 Mpc) \cite{hvsv,pb}. 
Consequently, the 
measurement of the diffuse gamma-ray
background imposes several constraints on the total amount of electromagnetic 
energy density injected and recycled by cascading to lower energies, yielding 
bounds on TD injection spectra \cite{sjsb}. It is also noteworthy that 
the expected high diffuse 
neutrino flux above 1 EeV sets additional bounds on the 
``top-down'' decay scenario \cite{slsc}.

The identity of the primary particle certainly deepens the mystery. 
Strictly speaking, a 100 EeV proton traverses a nearly straight line 
through the galactic or extragalactic magnetic fields and yet no 
compelling local sources have been identified 
within the expected scattering cone of the highest energy event \cite{es}. 
Furthermore, the atmospheric shower profile is reported by Fly's Eye \cite{FE}
to roughly fit that of a proton, or a gamma-ray that peaks somewhat late 
in the atmosphere \cite{hvsv}. It is also reported that a medium mass nucleus 
seems to fit this 
shower profile. 

Heavy nuclei have their own merits. 
They can be accelerated to higher terminal energies because of their larger 
charge. Additionally, a heavy nucleus
would have a larger scattering cone since they can be deflected 
more by the magnetic fields on the way to the observer. 
However, there are still 
other problems left in 
interpreting all the ``super-GZK'' events as heavy nuclei: It has long 
been known that above 200 EeV nuclei 
should be photodissociated by the MBR within a few Mpc \cite{puget}, and there 
are few nucleus-emitting-sources  in the neighborhood of the Earth.
More speculative primaries have also been  proposed, such as 
dust grains \cite{dg} or even neutrinos \cite{hvsv}. 
On the one hand, the grains turn out to be unstable with respect to the 
development of a fracture.
The path length up to the first break-up turns out to 
be much less than the characteristic size of the Milky 
Way \cite{berezinsky-book}. 
In addition, the depth of shower maxima registered by the Haverah Park and 
Volcano Ranch experiments do not correspond to the expected picture of showers 
initiated by relativistic specks of dust \cite{linsley}. Neutrinos, on the 
other hand, can traverse unscathed through the primeval radiation, but they 
would  predominantly  give rise to deeply penetrating showers in the 
atmosphere, again, in disagreement with observations. 
Putting all these together one can na\"{\i}vely say that aside from some 
nucleus species, the broad phylum of 
particles does not seem to fit the cascade development of the highest energy 
event.

All in all, the determination of the chemical composition appears crucial 
in discriminating between the possible origin of the highest energy 
cosmic rays. In light of this, we decided 
to re-examine  the  shower 
profile reported by the Fly's Eye experiment. In this letter we  
present an extension of the previous analysis \cite{hvsv} studying 
the effects of different atmospheric--cascade--models and parameter values.

The Fly's Eye observes an air shower as a nitrogen fluorescence light source 
which moves at the speed of light along the path of a high energy particle 
traversing the atmosphere \cite{NIM}. In other words, it directly detects the 
longitudinal 
development of the air cascade. The simulation of the shower evolution  
depends sensitively on the first 
few interactions, necessarily related to the quality of our ``understanding'' 
of hadronic collisions. Soft multiparticle production with small 
transverse momenta with respect to the collision axis is a dominant 
feature of most events in high energy hadronic interaction. 
Despite the fact that strict calculations based on ordinary QCD perturbation 
theory are 
not feasible, some phenomenological approaches successfully take into account
 the main 
properties of soft diffractive processes \cite{dpm,qgs,venus}. In these
models, the interactions are no longer described by single
particle exchange, but by highly complicated collective modes
known as Reggeons \cite{regge-gribov}. 
The slow growth of the
cross section with the centre of mass energy requires a dominant
contribution of a special Reggeon, the Pomeron \cite{pomeron}. As
a result of the diffractive interactions, constitutents get excited
and produce particles, which, in the fragmentation phase, are
modelled by strings \cite{sj}. The physical picture at 
super-high energies is quite different.
Interactions are not fully described within the framework of standard
Gribov-Regge theory \cite{regge-gribov}. At these extremely high energies,
apart from the usual excitations of the participants,
hard parton-parton collisions (between the beam and target hadrons) take
place. A suitable scenario for the joint description of both, soft and
semi-hard hadronic physics has yet to see the light of day.
The well known codes {\sc qgsjet} \cite{qgsjet}, {\sc sibyll} 
\cite{sibyll}, and {\sc dpmjet} \cite{ranft} represent three of the best 
simulation tools to model hadronic interactions at the highest 
energies. Let us now concentrate on the first two programs, and a 
discussion on {\sc dpmjet} will 
follow.\footnote{Improvements 
in the hadronic models are also under
way. {\sc nexus}, a joint enterprise by the authors
of {\sc venus} \cite{venus} and {\sc qgsjet} is expected to become
available soon. The preliminary {\sc nexus} skeleton has already been
described in \cite{nexus}.} 

The underlying idea behind  {\sc sibyll} is that the increase in
the cross section is driven by the production of
minijets \cite{sibyll}. The probability distribution for obtaining $N$
jet pairs (with $p_{\rm T}^{\rm jet}\,>\,p_{\rm T}^{\rm min}$, being
$p_{\rm T}^{\rm min}$ a sharp threshold on the transverse momentum
below which hard interactions are neglected)
in a collision  at energy $\sqrt{s}$ is computed regarding elastic $pp$ or
$p\bar{p}$ scattering as a difractive shadow scattering associated
with inelastic processes. The algorithms are
tuned to reproduce the central and fragmentation regions data
up to $p\bar{p}$ collider energies,
and with no further adjustments they are extrapolated several orders of 
magnitude.
On the other hand, in {\sc qgsjet} the theory is formulated entirely in
terms of Pomeron exchanges. The basic idea is to replace the soft Pomeron
by a so-called ``semihard Pomeron'', which is defined to be an ordinary soft
Pomeron with the middle piece replaced by a QCD parton ladder.
Thus, minijets will emerge as a part of the ``semihard Pomeron'',
which is itself the controlling mechanism for the whole 
interaction \cite{qgsjet}.
The different approaches used in both codes to model the underlying physics
show clear differences in multiplicity predictions which increase with 
rising energy \cite{prdhi}. It is important to stress that although the 
differences seem to become washed out as the shower front gets closer to 
the ground, the footprints of the first hadronic collisions are still 
present in the longitudinal development. Consequently, the first 
generation of particles plays a paramount role in  determining the 
chemical composition of the ``super-GZK'' event reported by Fly's Eye.

In order to determine the progenitor of the highest energy event,
we carried out a Monte Carlo simulation of the atmospheric shower profile
using the {\sc aires} program (version 2.2.1) \cite{sergio}. The simulation 
includes interaction with both the Earth magnetic field and nuclei 
in the atmosphere. We have taken  into account  possible representative 
variations in the transport code,  
cross sections, particle multiplicity and hadronic models by using 
{\sc sibyll} and {\sc qgsjet} to generate interactions above 200 GeV.

We generated several sets of nucleus showers 
with the mass  equally spread in the {\sc aires}  range (up to 
$^{56}$Fe). In addition, air-showers of superheavy nuclei were 
simulated following the superposition model. This model assumes that 
an average shower produced by a nucleus with energy 
$E$ and mass number $A$ is indistinguishable from  a superposition of $A$ 
proton showers, each with energy $E/A$.
The particles were injected  at the top of the atmosphere (100 km.a.s.l) 
with zenith angle 43.9$^\circ$, and the magnetic field was set to reproduce 
that prevailing upon Fly's Eye experiment \cite{analia}.
All shower particles with
energies above the following thresholds were
tracked: 750 keV for gammas, 900 keV for electrons and positrons, 10
MeV for muons, 60 MeV for mesons and 120 MeV for nucleons and nuclei.
The results of these simulations 
were processed with the help of the {\sc aires} analysis programs. 

The data analysis is performed by means of a $\chi^2$ test \cite{pdg}. 
We assume that the set of measured values by Fly's Eye are 
uncorrelated, i.e, any depth measurement is independent of each other. 
Then, we make use of the 
quantity 
\begin{equation}
\chi^2 \equiv \sum_{j=1}^N \frac{|x_j - \alpha_j|^2}{\sigma_{x_j}^2},
\end{equation}
where $N$ is the total number of points in the analysis, 
$\sigma_{x_j}$ is the error on the $x_j$th coordinate, $x_j$ is the measured 
value of the coordinate, and $\alpha_j$ the (hypothetical) true value of the 
coordinate. Generally speaking, the simulated shower 
profile would give a good representation of the data if the $\chi^2$ 
is roughly equal to the number of data points, i.e., $\chi^2 \approx 12$. 
For a gamma-ray primary 
we obtain $\chi^2_{\rm SIBYLL} = 1235.82$, and $\chi^2_{\rm QGSJET} = 1175.39$.
The results for different primary nuclei are shown in Fig. 1.  

Definite conclusions on the chemical composition cannot be reached, 
mainly because of large fluctuations from model to model. 
Without belaboring the point, the main differences between the 
hadronic-event-generators can be ascribed to two different parameters: 
the inelastic cross section, and the energy fraction of the parent particle 
converted into the production of secondary particles, the so-called 
``inelasticity''. In order to match collider data, the brutal truncation 
of the soft hadronic processes ({\sc sibyll}) enforces interactions with a 
large average fraction of the energy going into the leading 
baryon (elasticity) 
and a small fraction into secondaries (inelasticity) than those 
governed by {\sc qgsjet}. As a consequence, {\sc sibyll}'s showers require 
more generations of particles undergoing hadronic collisions, yielding
a delay in the electromagnetic shower development (strongly correlated 
with $\pi^0$ decays). However, both the cross section,
and the inelasticity seem to affect the shower development in a 
similar fashion since increasing inelasticity as well as cross section lead 
to faster cascades. Therefore, primaries with very different combination of 
cross section and inelasticity are able to describe the given longitudinal 
development.  In particular, the large elasticity of {\sc sibyll} yields a 
minimum on the $\chi^2$ above the {\sc aires} nucleus range. 
Here, the reader should keep in mind the crudeness of the superposition model 
for air shower simulations. In the 
simulations with 
{\sc qgsjet}+{\sc aires}, the $\chi^2$ reaches a 
minimum at $^{41}$Ca, leading to a plain representation of the shower 
profile (see Fig. 2).

The total error in the energy determination (systematic and
statistical uncertainties added in quadrature) reported by Fly's Eye is
93 EeV \cite{FE}. The next step in the analysis is to check whether the 
above results could be improved for any particular energy (within the error) 
by forcing the primary to be a proton. To this end we generate several sets 
of showers, varying the primary energy within the error box of the event.
The $\chi^2$  always remains very large. For example, in Fig. 1 we show 
the result (as a star) obtained for the upper limit of 
the energy for comparison. The last exercise is to 
analyze the sensitivity of the shower simulations. To handle 
shower-to-shower fluctuations we separately analyze each shower 
initiated by a proton: after 50 showers all the $\chi^2$s are above 20. 
Putting all this together, the point to be made is this: Even though the 
primary chemical 
composition remains hidden by the hadronic interaction model, it is evident 
that in both cases the shower is completely {\bf inconsistent} with a proton, 
or a gamma-ray primary.

At this stage, it is important to stress that although the latest 
version of {\sc dpmjet} was improved 
for  operation up to primary energies of $10^{21}$ eV (per nucleon in the lab. 
frame) \cite{ranft}, it has not yet been effectively 
implemented into the standard transport codes  \cite{talk}. However, 
one does not expect strong deviations in the longitudinal development if the 
hadronic collisions are modelled with {\sc dpmjet} \cite{knapp,knapp2}; the 
results for the $\chi^2$ should lie somewhere between those of 
{\sc aires}+{\sc sibyll} and {\sc aires}+{\sc qgsjet} (see, 
in particular Fig. 5 of Ref. \cite{knapp}).

Let us end  by reviewing the current observational status,
and discussing the nature of the highest energy CRs.  

\begin{itemize}

\item Up to now, the AGASA experiment reported 7 events above 100 EeV. 
Large shower to shower fluctuations make it difficult to get a definite 
dependence of composition from muon density at 1000m from the shower core 
(see Fig. 3 of Ref. \cite{inoue}). The situation is further complicated 
 if we take into account that not only two mass 
components (iron-proton) are allowed, but that many $A$-values could be 
present. There is no candidate shower of gamma-ray primary 
in this sample \cite{inoue}.  

\item The density profile of the highest energy event detected by 
the Yakutsk experiment is almost dominated by muons, which argues 
against proton/gamma-ray initiation \cite{yakutsk}. 

\item The  atmospheric profile of the highest energy cosmic ray is 
inconsistent with that initiated by a proton or a gamma-ray. 
Moreover, an analysis of the frequency 
distribution of the depth maximum values recorded by Fly's Eye, indicates 
that in the energy range 3 - 10 EeV, there is a significant fraction of 
nuclei with charge greater than one \cite{ww}.  

\item Very recently, the Haverah Park experiment reported data from 
inclined showers \cite{ave-etal}.  At the ``super-GZK'' energies, the 
resulting spectrum is consistent with nucleus primaries (see 
Fig. 2 of Ref. \cite{ave-etal}). We stress once more that above the 
photodisintegration threshold different type of debris may reach the 
atmosphere; consequently, the primary mixture is far from just protons 
and iron nuclei. This sample also states evidence against the photon 
hypothesis.

\end{itemize}

Whatever the source(s) of the highest energy cosmic rays, because of their 
interactions with magnetic fields, and also with the universal diffuse 
radiation backgrounds permeating the universe, 
the energy spectra, composition and arrival directions are  affected by 
propagation. In particular, in the case of heavy nuclei, the composition of 
the arriving particles may differ substantially from that of the source.  
For instance, an iron nucleus which is injected into the 
intergalactic medium with a Lorentz factor of $\gamma_0 = 4 \times 10^9$, 
after propagating $\approx 100$ Mpc will end up with 
$A \approx 30$.\footnote{It should be noted, that the 
photodisintegration history of the nucleus was performed using the 
single/double-nucleon emission energy threshold 
of ref. \cite{ss}. The value of $A$ is thus slightly different 
to that of Ref. \cite{luis-esteban}, because the increase of the threshold 
energy lengthens the propagation distance.} Such 
metamorphosis is the result of two basic interactions: photodisintegration 
and pair creation.  The latter reduces 
the Lorentz factor significantly during propagation avoiding the 
complete disintegration of the nucleus.  
Stecker and Salamon have pointed out that a heavy nucleus 
emitted by the strong radio galaxy NGC315, which is 60 Mpc away and within 
the scattering cone of the highest energy event detected by AGASA, would 
have an energy cut off $\approx 130$ EeV, which may be within the 
uncertainty in the energy 
determination for this event \cite{ss}. Furthermore, it was 
already noted that a positively charged particle injected by the next-door 
galaxy M82 may have been deflected towards the arrival direction of the 
highest energy event \cite{ngc}. Note that although M82 is not a strong radio 
galaxy, it has been described as the archetypal starburst galaxy 
(supernova rate as high as 0.2 -- 0.3 yr$^{-1}$) and as a prototype of 
superwind galaxies. In this picture, the central region of the galaxy 
harbors a population  of heavy ions (mainly  produced in supernova explotions) 
that could be re-accelerated to ultra high energies at the 
terminal shocks of galactic superwinds generated by the starburst. 
It is noteworthy that M82 could also be responsible for 
the ``super-GZK'' Yakutsk event.  In addition, it is 
interesting to point out that the regular component of 
the galactic magnetic field acts as a giant lens for nuclei where  
$E/Z \leq 50$ EeV, yielding significant deflections of CR 
trajectories \cite{toes}. 

Finally, we would like to stress another remarkable feature of nucleus 
primaries. It was already noted that nuclei heavier than iron could be 
accelerated to super-high energies \cite{neu3}. Moreover, 
the dominant mechanism for energy losses in the bath of the universal 
cosmic radiation is the photodisintegration process, which was studied 
in detail for the case of $^{197}$Au (see Fig. 2 of Ref. \cite{neu3}). 
In particular, the photodisintegration rate $R$ of a gold nucleus with Lorentz 
factor $\gamma_0 = 1 \times 10^{9}$ ($\gamma_0 = 2 \times 10^{9}$) is  
$R= 3.22 \times 10^{-15}$ s$^{-1}$ ($R= 1.55 \times 10^{-14}$ s$^{-1}$), and 
the energy loss time is defined by
\begin{equation}
\tau^{-1} \equiv \frac{1}{E} \frac{dE}{dt} = \frac{1}{\gamma} 
\frac{d\gamma}{dt} + \frac{R}{A}.
\end{equation}
The energy loss due to photopair production could be estimated to be roughly 
$Z^2/A$ times higher than that of a proton with the same Lorentz
factor, i.e., for $\gamma_0 \approx 10^{9}$, $\tau^{-1}_{e^+ e^-} \equiv
1/\gamma\,d\gamma/dt \approx 
2 \times 10^{-18}$ s$^{-1}$ \cite{e+e-}. For a gold nucleus then we get, 
$\tau \approx 6 \times 10^{16}$s  ($\tau \approx 1 \times 10^{16}$ s). 
Now, using the previous formulae for $\tau$, it is 
straightforward to show that a gold nucleus emitted with a 
Lorentz factor  \mbox{$\approx 10^{9}$,} after propagating 100 Mpc 
still has an energy above 100 EeV. Specifically, gold nuclei 
injected with 197 EeV, will mutate to 
$A \approx 166 - 168$ with an energy $\approx 167$ EeV, 
whereas for $E_0 = 394$ EeV, the surviving fragment would 
have $A=89-91$ with an energy $\approx$ 180 EeV. 
Note that both values are within 1 standard deviation of the 
highest energy event recorded by AGASA ($E$ = 210 EeV, whose accuracy 
on energy 
determination is 30\% \cite{agasa}). Furthermore, in this case the deflections 
of the CR trajectories would be enough to conceal the source location. 
The above valuation  together with our previous analysis suggest 
that elements heavier than iron are likely to generate some of 
the ``super-GZK'' events. It perhaps should be interesting to try to 
simulate  showers initiated by super-heavy nuclei considering 
the internal nuclear structure, specially when the hadronic interactions 
are modelled with {\sc sibyll}.

In summary, nucleus induced showers are in complete agreement with the 
experimental data reported by the groups of Fly's Eye, 
Yakutsk, AGASA and Haverah Park. Moreover, if the highest energy CRs are 
heavy nuclei, one cannot yet rule out that extragalactic/galactic 
magnetic fields could tangle up the particle paths, camouflaging the  
exact location of sources. The astrophysics parameters governing CR 
propagation are still 
somewhat uncertain, and better statistics on the arrival directions, 
energy spectra and composition are indispensable for  understanding the 
origin of the highest energy particles ocurring in nature. Nonetheless, 
it does not seem reasonable to invoke ``top-down'' models to explain the 
existing 
data since they would fail to explain the nucleus--component, 
certainly present at the end of the spectrum.

\section*{Acknowledgments}

Special thanks go to Gabriela Anchordoqui for carefully extracting the 
experimental data from Fig. 3  of Ref. \cite{FE}. The work was supported by
CONICET (Argentina), and the National Science Foundation.

\newpage

\begin{figure}
\begin{center}
\epsfig{file=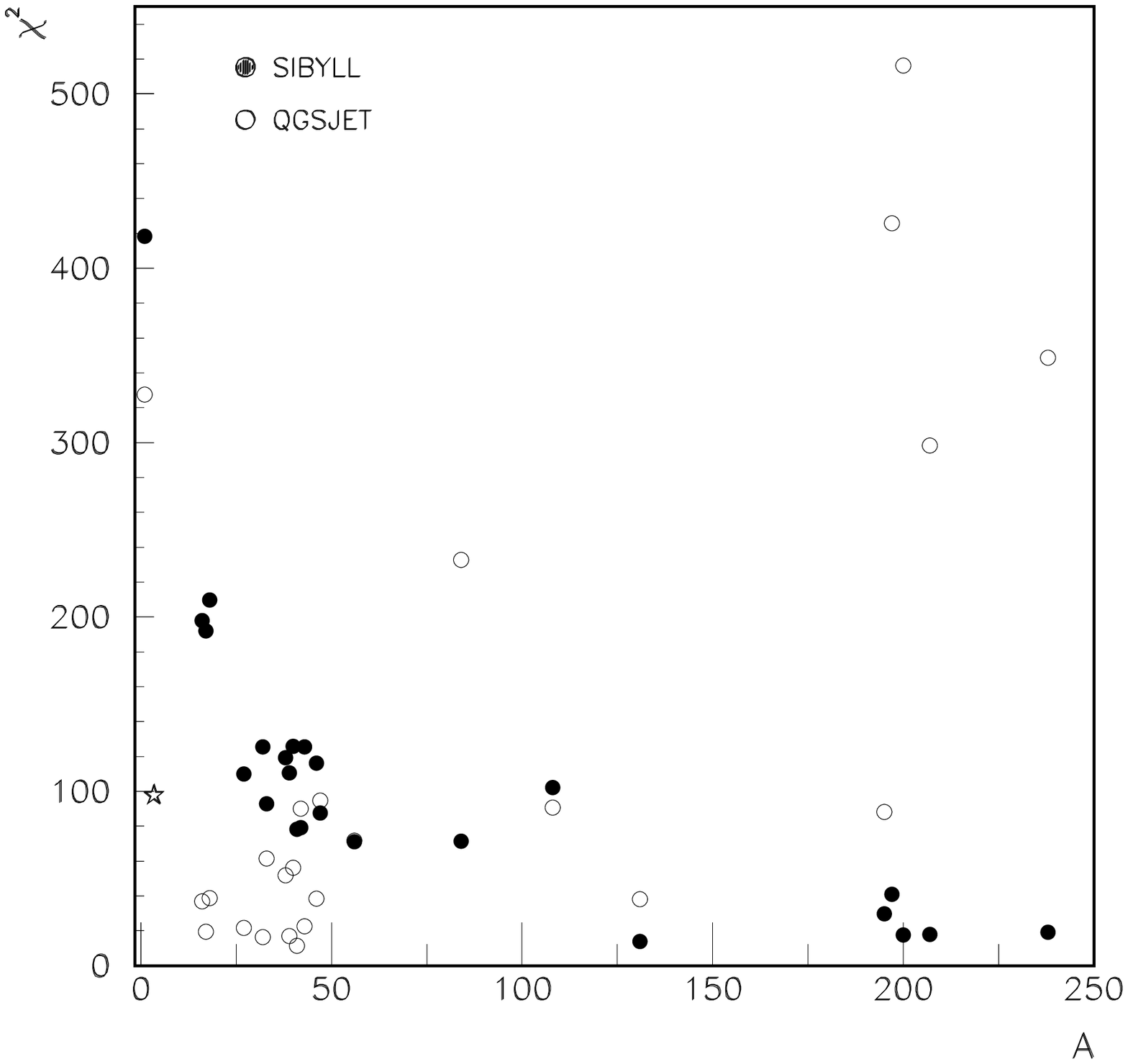,width=14.5cm,clip=} 
\caption{Behavior of $\chi^2$ as a function of number of nucleons $A$. The 
white star stands for the $\chi^2$ of a 355 EeV proton induced shower 
({\sc aires}+{\sc qgsjet}, see main text).
}
\end{center}
\end{figure}

\newpage

\begin{figure}
\begin{center}
\epsfig{file=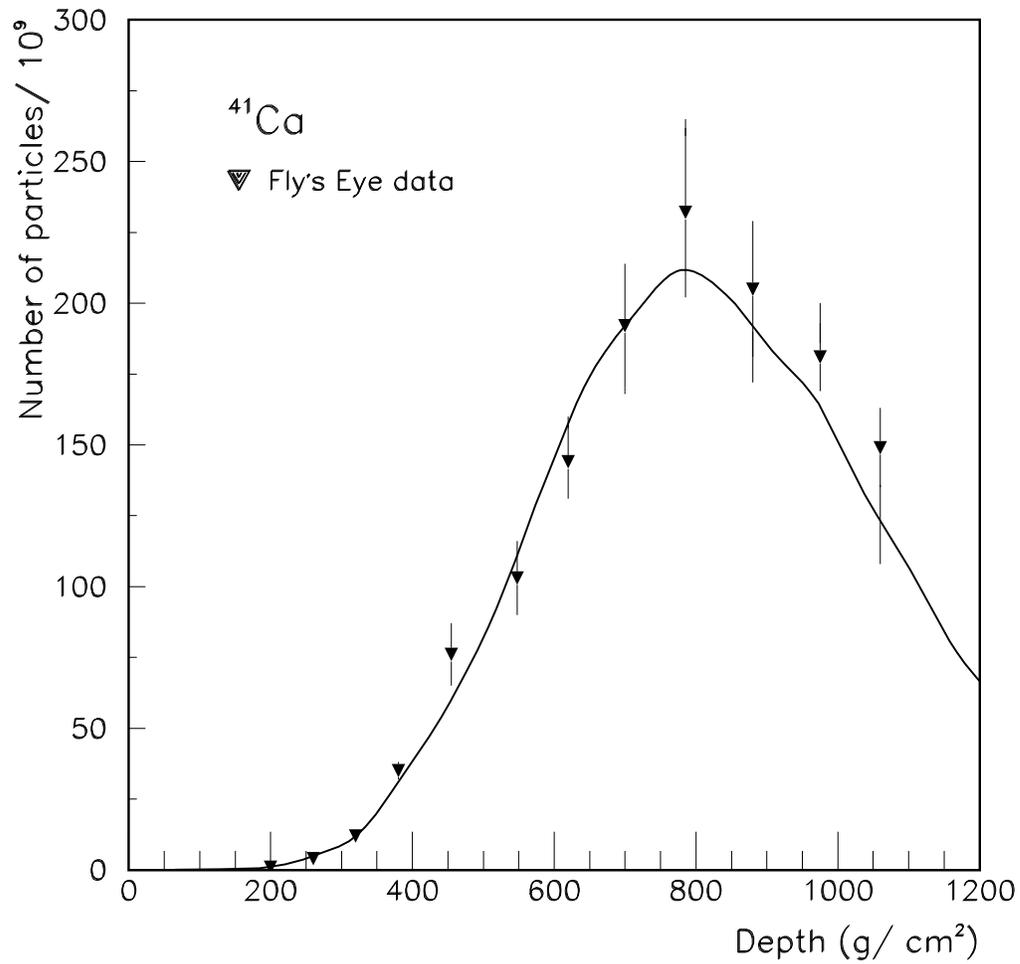,width=14.5cm,clip=} 
\caption{Atmospheric cascade development of a 300 EeV nucleus induced 
shower, superimposed over the Fly's Eye data.}
\end{center}
\end{figure}

\end{document}